\documentclass[
journal=jpclcd,
manuscript=letter, review
]{achemso}

\usepackage{graphicx}
\usepackage{dcolumn}
\usepackage{bm}
\usepackage{xcolor}
\usepackage{amsmath}
\usepackage{mathrsfs}

\usepackage[utf8]{inputenc}
\usepackage[T1]{fontenc}
\usepackage{etoolbox}
\usepackage{booktabs}
\usepackage{gensymb}
\usepackage[colorlinks=true,citecolor=blue,linkcolor=blue, urlcolor=blue]{hyperref}
\usepackage[font=footnotesize,labelfont=bf, textfont=normal]{caption}
\usepackage[version=3]{mhchem} 
\usepackage{dsfont}
\usepackage{pdfpages}

\newcommand{\braket}[1]{\langle #1 \rangle}

\author{Sara Angelico}
\affiliation{Department of Chemistry, Norwegian University of Science and Technology, NTNU, 7491 Trondheim, Norway}
\author{Eirik F. Kjønstad}
\affiliation{Department of Chemistry, Norwegian University of Science and Technology, NTNU, 7491 Trondheim, Norway}
\author{Henrik Koch}
\affiliation{Department of Chemistry, Norwegian University of Science and Technology, NTNU, 7491 Trondheim, Norway}
\email{henrik.koch@ntnu.no}

\title[]
  {Determining minimum energy conical intersections by enveloping the seam:
  exploring ground and excited state intersections in coupled cluster theory}

\begin{document}

\begin{tocentry}

   \includegraphics[width=\columnwidth]{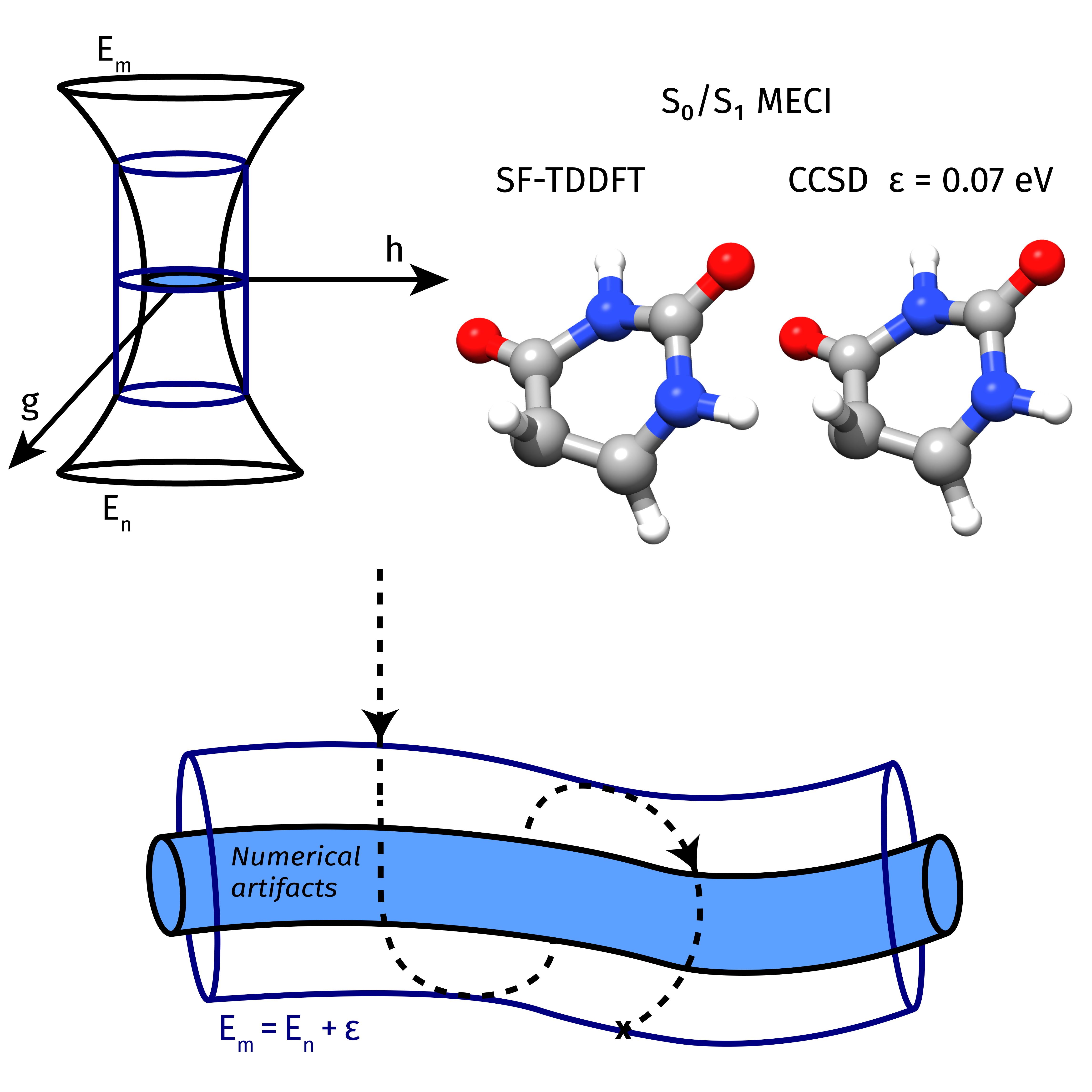}
    
\end{tocentry}

\begin{abstract}

Minimum energy conical intersections can be used to rationalize photochemical processes. In this Letter, we examine an
algorithm to locate these structures that does not require the 
evaluation of
nonadiabatic coupling vectors, showing that it minimizes the energy on hypersurfaces that envelop the intersection seam. 
By constraining the states to be separated by a 
small non-zero energy difference, the algorithm ensures that numerical artifacts and convergence problems of coupled cluster theory at conical intersections are not encountered during the optimization. 
In this way, we demonstrate for various systems that minimum energy conical intersections with the ground state are well described by the coupled cluster singles and doubles model, suggesting that coupled cluster theory may in some cases
provide a good description of relaxation to the ground state in nonadiabatic dynamics simulations.

\end{abstract}

Degeneracies between adiabatic states, or conical intersections, play a fundamental role in nonadiabatic dynamics. In these regions, two or more potential energy surfaces cross, allowing for the transfer of population between different states.
It is now widely recognized that conical intersections are widespread in molecular systems.\cite{yarkony1990characterization,yarkony1996diabolical,atchity1991potential,bernardi1990mechanism}
Given a molecule with $N$ internal degrees of freedom, the region of the configuration space where two same-symmetry states are degenerate has dimension $N-2$.\cite{von1993verhalten} This defines the crossing seam, which is assumed to be a Riemann manifold, that is smooth and everywhere differentiable.
In the vicinity of the intersection, the degeneracy is lifted linearly in two directions and the potential energy surfaces describe a double cone. The dimensionality of the crossing seam and the shape of the potential energy surfaces are referred to as the topology and topography of the conical intersection, respectively.\cite{matsika2021electronic}

Determining the location of these intersections is in general difficult, as they are points of accidental degeneracy.\citep{zhu2016non}
However, the rationalization of ultrafast photochemical processes often benefits from information about the critical points on the energy surfaces, for example, the minimum energy conical intersection (MECI) structures. 
Therefore, developing black box algorithms to determine MECIs is an area of continued focus.
The first automated numerical algorithm, developed by Yarkony and co-workers,~\cite{manaa1993intersection} is a second-order algorithm that makes use of the molecular gradient, Hessian, and nonadiabatic coupling vectors (or derivative couplings). Since then, the algorithm has been improved~\cite{yarkony2004marching} and other approaches have been proposed, such as gradient projection methods.\cite{BEARPARK1994269,sicilia2008new} While these methods are robust, they rely on derivative couplings, which are not widely available in electronic structure programs.
As a result, the focus has shifted toward developing algorithms that do not require derivative couplings. Examples include penalty function methods\cite{levine2008optimizing} and the branching plane updating method~\cite{maeda2010updated}. We refer to Ref.~\citenum{matsika2021electronic} and references therein for an extensive description of the existing algorithms.

The correct description of conical intersections can be challenging for electronic structure methods.\cite{matsika2021electronic} It requires a balanced treatment of the states involved, as well as the ability to describe the correct dimensionality of the crossing seam and topography of the surfaces. Multiconfigurational methods, such as complete active space self-consistent field (CASSCF\citep{roos1980complete}) and its variants with perturbation theory corrections (e.g., CASPT2\citep{andersson1992second}), can provide a balanced description of ground and excited states and are therefore often used to describe nonadiabatic processes. Single reference methods, on the other hand, can offer a balanced treatment of excited states, but accurately describing ground state intersections remains particularly challenging. This is true for time-dependent density functional theory (TD-DFT); however, various modifications have been proposed to mitigate these complications.\cite{teh2019simplest,li2014configuration,shu2017dual,calio2024minimum} 

Coupled cluster theory is also known to have serious problems with the description of the topology and topography of conical intersections among excited states,\cite{hattig2005structure,kohn2007can} where TD-DFT is known to work in the Tamm-Dancoff approximation.\cite{levine2006conical} The intersection problem among excited states in coupled cluster theory was recently resolved with the introduction of similarity constrained coupled cluster (SCC) theory.\cite{kjonstad2017resolving,kjonstad2019orbital}
 The SCC framework has already been successfully applied by Kjønstad et al.~\cite{kjonstad2024unexpected,kjonstad2024coupled} to nonadiabatic dynamics, and an efficient implementation of molecular gradients and nonadiabatic couplings has been developed.
 However, intersections with the ground state remain an open problem. 
 As we will show, the algorithm we analyze below can avoid the unphysical artifacts in coupled cluster theory and furthermore locate ground state conical intersection structures. As a result, it may provide insight into a possible solution to the ground state intersection problem in coupled cluster theory.~\cite{rossi2024generalized}

\begin{figure}[!htbp]
    \includegraphics[width=0.88\columnwidth]{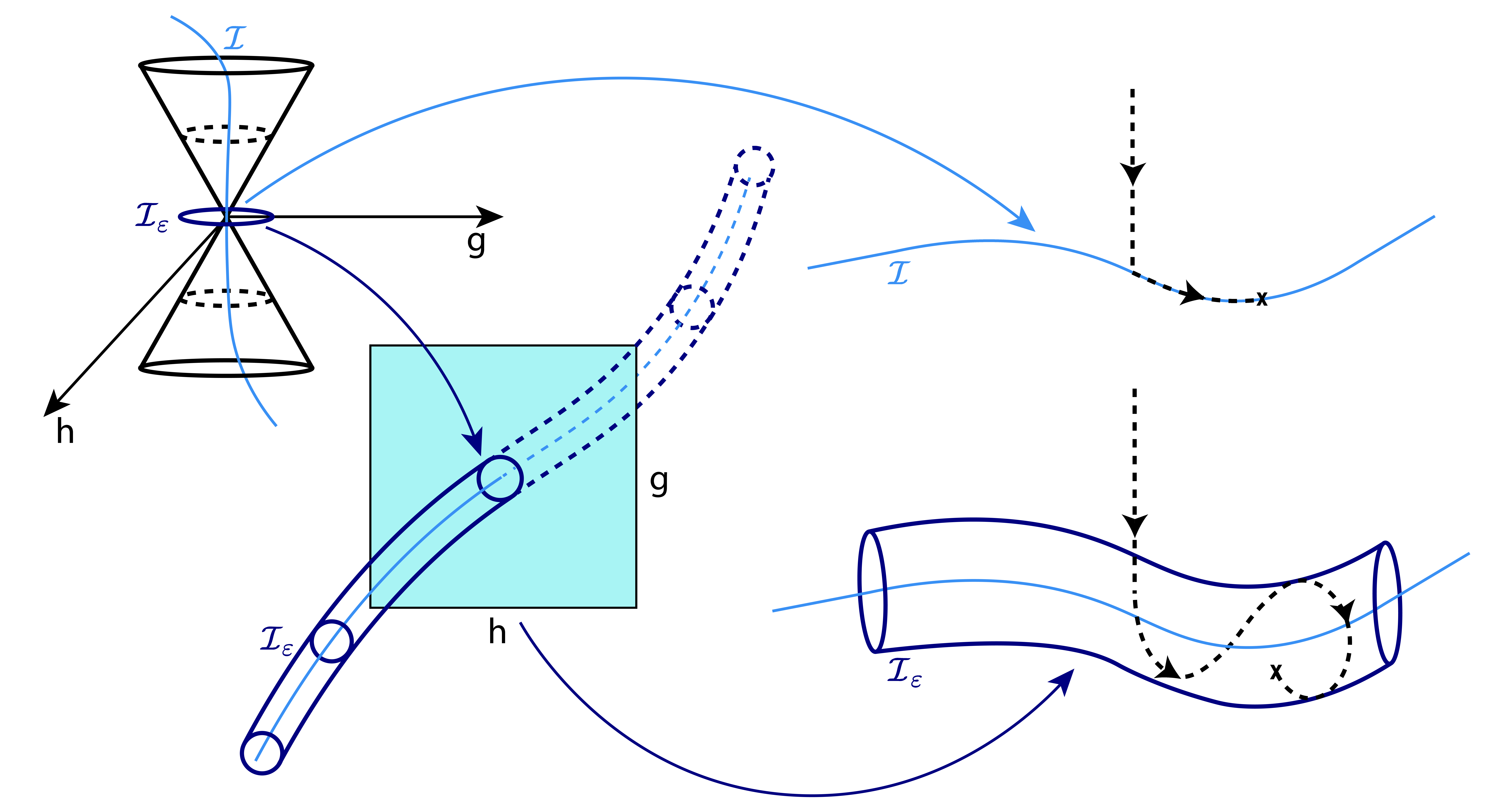}
    \caption{\footnotesize
    The intersecting energy surfaces describe a double cone in the g-h plane (top left). 
    The energy difference is $\varepsilon$ along an ellipse identified by $\mathcal{I}_\varepsilon$. 
    The crossing seam $\mathcal{I}$ is orthogonal to the plane. Moving along the seam, $\mathcal{I}_\varepsilon$ describes a tube (center). 
    In both the gradient projection method\cite{BEARPARK1994269} (top right) and the tube algorithm (bottom right), $\mathcal{I}$ or $\mathcal{I}_\varepsilon$ are first reached from the starting guess. Then, the energy of the upper state is minimized by moving along the seam or along the surface of the tube, respectively.}
    \label{fig:algorithms-comparison}
\end{figure}

We start by briefly summarizing the gradient projection method.~\citep{BEARPARK1994269} 
We define the crossing seam $\mathcal{I}$ as the set of geometries where two states $m$ and $n$ are degenerate,
\begin{equation}
    \mathcal{I} = \{ \mathbf{R} : E_m (\mathbf{R}) - E_n (\mathbf{R}) = 0 \} .
\end{equation}
The orthogonal complement to 
a crossing point
is the space where the degeneracy is lifted, usually called the branching plane.
The branching plane is spanned by two vectors, $\boldsymbol{g}_{nm} = \boldsymbol{\nabla} (E_n - E_m)$ and $\boldsymbol{h}_{nm} = \braket{\psi_n | \boldsymbol{\nabla} \psi_m}(E_m - E_n)$.
In the algorithm developed by Bearpark et al.~\cite{BEARPARK1994269}, the energy of the upper state is minimized along the crossing seam. This is obtained by minimizing the gradient
\begin{equation}
\label{eq:bearpark}
    \mathbf{G}_{nm} = \mathcal{P}_{nm} \boldsymbol{\nabla}E_m + 2(E_m - E_n) \frac{\boldsymbol{g}_{nm}}{||\boldsymbol{g}_{nm}||},
\end{equation}
where $\mathcal{P}_{nm}$ is a projector along the crossing seam
or, equivalently, on the orthogonal complement to the branching plane,
\begin{align}
 \mathcal{P}_{nm} = \mathds{1} - \boldsymbol{g}_{nm}(\boldsymbol{g}_{nm})^T - \boldsymbol{h}_{nm}(\boldsymbol{h}_{nm})^T.   
\end{align}
We may view $\mathbf{G}_{nm}$ in eq.~\ref{eq:bearpark} as being composed of two orthogonal contributions each serving different purposes. The first term minimizes the energy of the upper state along the crossing seam, whereas the second term minimizes the energy difference between the two states.
The minimization then becomes a two-step process. Initially, the states are far apart and the second term dominates; then, after reaching the crossing seam, the first term will direct the way to the minimum energy conical intersection. A schematic illustration of this process is shown in Fig.~\ref{fig:algorithms-comparison}.

A similar algorithm has been proposed to avoid the degeneracy by applying a small constant shift $\varepsilon$ to one of the states involved.~\citep{harabuchi2019exploring}
Adopting an analogous approach, we focus on the region of the internal coordinate space where the energy difference between the two states is constant. 
We define for a given $\varepsilon$ the $(N-1)$ dimensional space
\begin{equation}
    \mathcal{I}_{\varepsilon} = \{ \mathbf{R} : E_m (\mathbf{R}) - E_n (\mathbf{R}) = \varepsilon \} .
\end{equation} 
In the vicinity of the conical intersection, we know the degeneracy is lifted in two directions only, see Fig.~\ref{fig:algorithms-comparison}. 
In the branching plane, $\mathcal{I}_\varepsilon$ describes an ellipse, depending on the shape of the double cone. 
When moving along the seam, which is orthogonal to the branching plane, $\mathcal{I}_\varepsilon$ describes a tube. 
Note that this is not the case for linear intersections, although $\mathcal{I}_\varepsilon$ still encloses the seam in the sense that $\mathcal{I}$ can only be reached by passing through $\mathcal{I}_\varepsilon$. Such intersections can occur due to symmetry or due to an incorrect description by the electronic structure method (e.g., for TD-DFT\citep{levine2006conical}).
For sufficiently small values of $\varepsilon$, the tube folds around the crossing seam, and, in the limit of $\varepsilon\rightarrow 0$, the tube collapses to the crossing seam. By locating the minimum energy geometry on the tube
$\mathcal{I}_\varepsilon$, we obtain an accurate approximation to the minimum energy conical intersection for sufficiently small $\varepsilon$. We will refer to the approach as the tube algorithm.
The minimum energy conical intersection on $\mathcal{I}_\varepsilon$ can be determined by locating a zero
of the modified gradient 
\begin{equation}
\label{eq:our-algorithm}
    \mathbf{G}_{nm}^{\varepsilon} = \mathcal{P}_{nm}^\varepsilon\boldsymbol{\nabla}E_m + 2 (E_m - E_n - \varepsilon) \frac{\boldsymbol{g}_{nm}^\varepsilon}{||\boldsymbol{g}_{nm}^\varepsilon||} ,
\end{equation}
where we have defined $\boldsymbol{g}_{nm}^\varepsilon = \boldsymbol{\nabla}(E_m - E_n - \varepsilon) = \boldsymbol{g}_{nm}$. 
In analogy with the previous case, 
$\mathcal{P}_{nm}^\varepsilon$ is a projector on the orthogonal complement of $\mathcal{I}_\varepsilon$, which is spanned by $\boldsymbol{g}_{nm}$. This can be seen by noting that $\mathcal{I}_\varepsilon$ defines a level set of the function $(E_m - E_n)$, to which $\boldsymbol{g}_{nm}$ is orthogonal.~\cite{herbert2008analysis} As a result, the projected gradient minimizes the energy on the surface of the tube $\mathcal{I}_\varepsilon$. We have that
\begin{equation}
    \mathcal{P}_{nm}^\varepsilon = \mathds{1} - \boldsymbol{g}_{nm}\boldsymbol{g}_{nm}^T . 
\end{equation}
Note that the gradient function in eq.~\ref{eq:our-algorithm} coincides with the one applied in Ref.~\citenum{harabuchi2019exploring}, where $\boldsymbol{g}_{nm}$ was similarly projected out. From the above, we see that this choice actually corresponds to a minimization on the hypersurface $\mathcal{I}_\varepsilon$. 
Note also that evaluating the gradient
in eq.~\ref{eq:our-algorithm} only requires the $\boldsymbol{g}_{nm}$ vector, which can be obtained at different levels of theory by using analytical implementations or numerical differentiation of the potential energy surfaces. 

The localization of the approximate MECI consists in two main steps. 
First, when the potential energy surfaces are widely separated in energy, the second term in eq.~\ref{eq:our-algorithm} dominates, and this leads the optimization towards the isosurface $\mathcal{I}_\varepsilon$. Once the surface is reached, the first term of eq.~\ref{eq:our-algorithm} becomes dominant, and the algorithm moves along the surface of the tube $\mathcal{I}_\varepsilon$ in order to minimize the energy of the upper state. An illustration of this procedure is given in Fig.~\ref{fig:algorithms-comparison}. Since the algorithm
locates minimum energy geometries on the surface of the tube $\mathcal{I}_\varepsilon$, we will refer to the converged molecular structures as $\varepsilon$-MECIs.

We should point out that since the optimization identifies a geometry where the energy difference between the states is $\varepsilon$, the algorithm can in principle converge to an avoided crossing. This limitation is common in algorithms that locate MECIs. 
The identification of a real conical intersection can be confirmed by observation of the geometric phase effect in the electronic wave function along a path enclosing the identified intersection.\citep{longuet1975intersection} Moreover, avoided crossing geometries still represent meaningful areas of the nuclear configuration space, since they can also mediate transfer of population between the potential energy surfaces.\cite{levine2008optimizing}

Finally, while the algorithm is suitable for any electronic structure method, it is particularly useful for coupled cluster methods. Here, we can avoid regions with numerical artifacts and convergence problems\cite{kjonstad2024understanding} during the optimization procedure when $\varepsilon$ is chosen appropriately. In this case, it is particularly useful to adopt a stepwise procedure, where a 
large value of $\varepsilon$ is first used to locate an initial geometry. 
This geometry
is then used as the starting point for a second optimization with a smaller $\varepsilon$. For a given optimization step, the convergence properties of the algorithm do not differ significantly from that of the gradient projection method (see Supporting Information and Table~\ref{tab:convergence}).

\begin{table}[ht!]
\footnotesize
\centering
\begin{tabular}{lcccccc}
\hline
\hline
& $n_{\mathrm{iter}}$ \\ 
\hline
Gradient projection & 51 \\
Tube ($\varepsilon$=0.27 eV) & 30 \\
Tube ($\varepsilon$=0.027 eV) & 47 \\
\hline
\hline
\end{tabular}
\caption{Number of iterations $n_{\mathrm{iter}}$ required to converge the minimum energy conical intersections between S$_1$ and S$_2$ in uracil using the gradient projection and tube algorithm. The optimizations are for CCSD using the cc-pVDZ basis. The calculation with the smaller $\varepsilon$ was restarted from the calculation with the larger $\varepsilon$.}
\label{tab:convergence}
\end{table}

To assess the performance of the algorithm, we first determine $\varepsilon$-MECIs  between $\mathrm{S}_1$ and $\mathrm{S}_2$ in uracil using equation of motion  
coupled cluster singles and doubles (CCSD)~\citep{koch1990coupled,stanton1993equation} and its similarity constrained variant (SCCSD).~\citep{kjonstad2019orbital}
All calculations have been performed using a development branch of the $e^T$ program,\cite{folkestad20201} where this algorithm and analytical molecular gradients are implemented.\cite{schnack2022efficient,kjonstad2024coupled} In the MECI optimizations, the existing $e^T$ implementation for geometry optimization is used. Here, the gradient in eq.~\ref{eq:our-algorithm} is used in a BFGS algorithm that performs the optimization using redundant internal coordinates.\citep{bakken2002efficient} All structures presented are available in a separate repository.~\citep{zenodo}

\begin{figure}[!htbp]
    \includegraphics[width=\columnwidth]{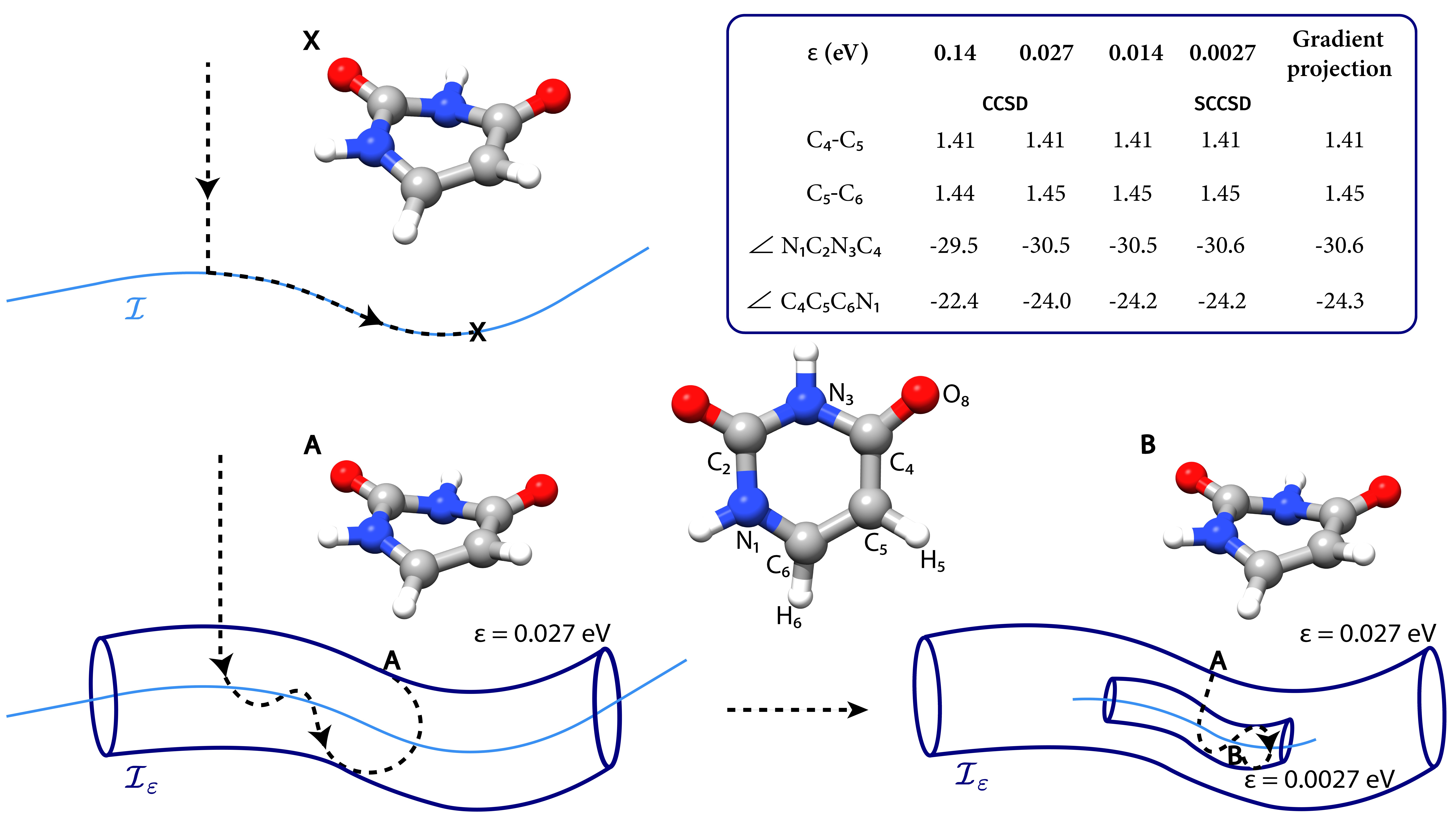}
    \caption{\footnotesize Illustration of a step-wise optimization of a S$_1$/S$_2$ $\varepsilon$-MECI for uracil, using CCSD ($\varepsilon$=0.14, 0.027 eV) and SCCSD ($\varepsilon$=0.014, 0.0027 eV), and comparison with the MECI determined using the gradient projection method (with SCCSD). The basis set is cc-pVDZ. Bond lengths are expressed in \AA~and angles in degrees ($\degree$). The optimized geometries from the gradient projection method, the tube algorithm with $\varepsilon=0.027$ eV and $\varepsilon=0.0027$ eV are shown in the figure as \textbf{X}, \textbf{A} and \textbf{B}, respectively.
}

    \label{fig:shrinking-tube}
\end{figure}

Converged S$_1$/S$_2$ MECIs for uracil are shown in Fig.~\ref{fig:shrinking-tube}.
In our optimization, we adopt a stepwise procedure  
where we start by using CCSD and $\varepsilon=0.14$ eV.
We employ SCCSD for $\varepsilon \leq 0.014$ eV
to avoid numerical artifacts during the optimization. Note that the SCCSD corrections in the energy and in the optimal geometry for $\varepsilon = 0.027$ eV 
are negligible (see Supporting Information). 
We compare the converged $\varepsilon$-MECI geometries with 
the one obtained by applying the gradient projection algorithm for SCCSD.
As can be seen from Fig.~\ref{fig:shrinking-tube}, by decreasing the value of $\varepsilon$, the $\varepsilon$-MECI converges to the MECI determined by using the gradient projection algorithm. 
Moreover, the $\varepsilon$-MECI determined with CCSD for $\varepsilon$=0.027 eV is 
highly similar to the MECI structure determined with SCCSD and the gradient projection algorithm.

\begin{figure}[!htbp]
    \includegraphics[width=0.5\columnwidth]{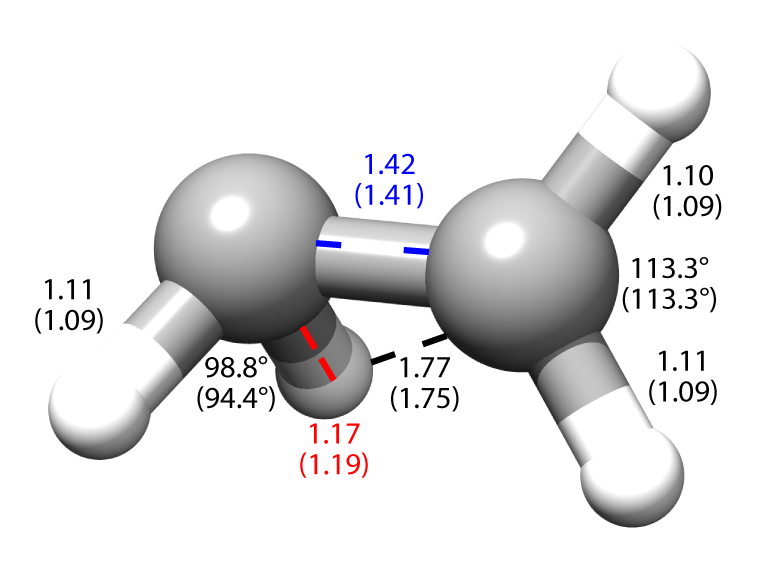}
    \caption{\footnotesize The S$_0$/S$_1$ pyramidalized MECI for ethylene. Internal coordinates for the CCSD $\varepsilon$-MECI are shown for $\varepsilon$= 0.27 eV, basis set aug-cc-pVDZ. Reference values from Ref. \citenum{ben2000photodynamics} determined with 2SA-CASSCF(4/7)/aug-cc-pVDZ are reported in parenthesis. Bond lengths are expressed in \AA~and angles in degrees ($\degree$). }
    \label{fig:ethylene-comparison}
\end{figure}

\begin{figure}[!htbp]
    \includegraphics[width=\columnwidth]{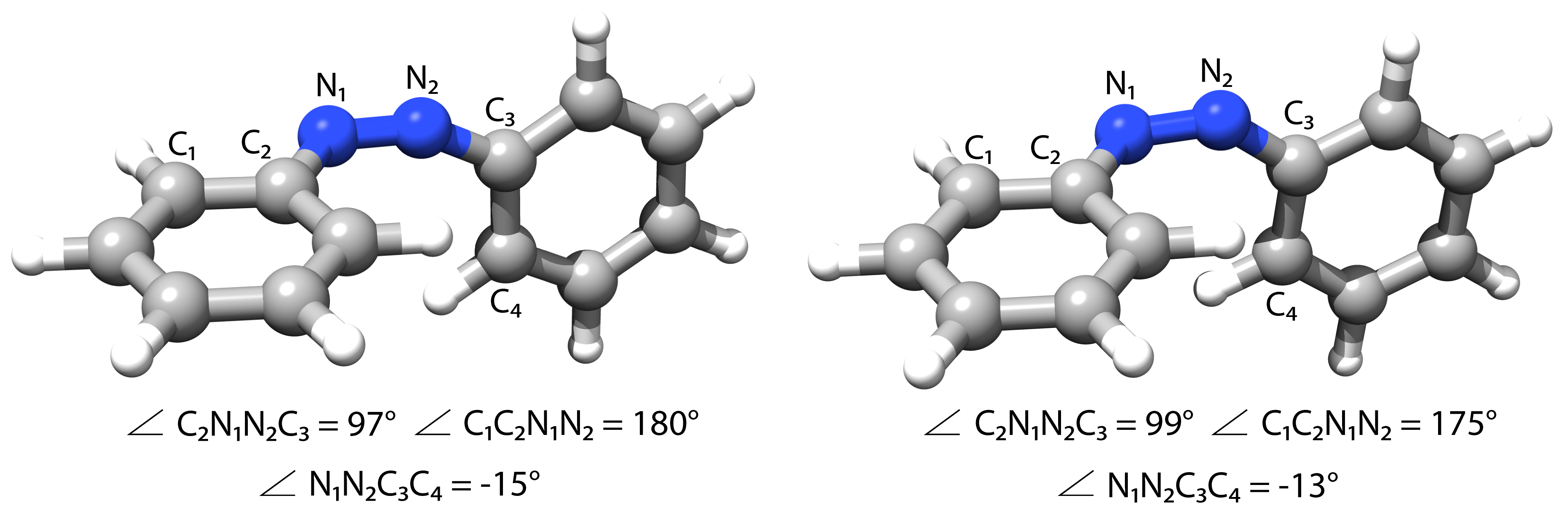}
    \caption{\footnotesize S$_0$/S$_1$ MECI for azobenzene with 5SA-CASSCF(6/6)/6-31G from Ref. \citenum{yu2015probing} (left) and $\varepsilon$-MECI with CCSD/6-31G (right). $\varepsilon$= 0.20 eV.
    Further comparison of internal coordinates is provided in the Supporting Information.} 
    \label{fig:azobenzene-comparison}
\end{figure}

\begin{figure}[!htbp]
    \includegraphics[width=0.9\columnwidth]{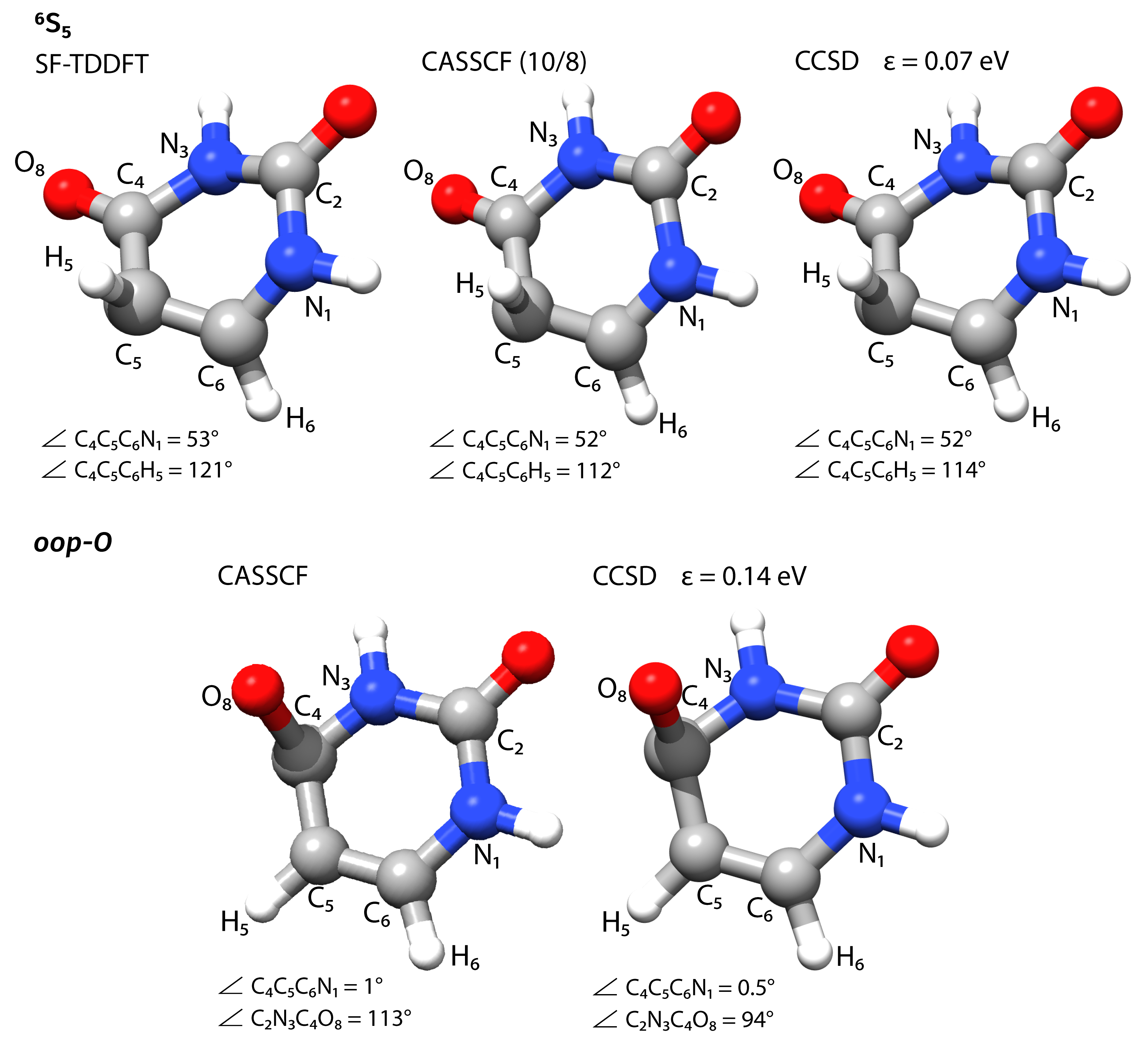}
    \caption{\footnotesize S$_0$/S$_1$ MECIs for uracil. The $^6$S$_5$ structures were determined at the SF-TDDFT/6-31+G(d,p) level in Ref. \citenum{zhang2014excited} (top left), CASSCF (10/8)/6-31G* level in Ref. \citenum{nachtigallova2011nonadiabatic} (top center) and CCSD/cc-pVDZ $\varepsilon$=0.07 eV (top right). The \textit{oop}-O structures were determined at the CASSCF(10/8)/6-31G* level in Ref. \citenum{nachtigallova2011nonadiabatic} (bottom left) and CCSD/cc-pVDZ $\varepsilon$=0.14 eV (bottom right).
    Further comparison of internal coordinates is provided in the Supporting Information.}
    \label{fig:uracil-comparison}
\end{figure}

The tube algorithm can also be applied to determine ground state MECIs. We now apply the tube algorithm to $\mathrm{S}_0$/$\mathrm{S}_1$ conical intersections for ethylene, azobenzene and uracil using CCSD. The molecular structures, together with reference geometries from the literature and numbering of the atoms, are reported in Figs.~\ref{fig:ethylene-comparison}, ~\ref{fig:azobenzene-comparison} and~\ref{fig:uracil-comparison}. In all cases, we present here results for the smallest value of $\varepsilon$ where the algorithm was able to converge. Additional results with different values of $\varepsilon$ are provided in the Supporting Information. 

For ethylene, we locate the pyramidalized S$_0$/S$_1$ MECI, with $\varepsilon$= 0.27 eV using the aug-cc-pVDZ basis set. As can be observed in Fig.~\ref{fig:ethylene-comparison}, the CCSD $\varepsilon$-MECI structure reproduces the state-averaged CASSCF (SA-CASSCF) results reported in Ref~\citenum{ben2000photodynamics}. In particular, the \ce{C-C} and \ce{C-H} bond lengths only differ from the reference geometry by 0.02 \AA~at most. The pyramidalization of the C atom is also well described, as can be seen by the distance between the carbon atoms and their non-adjacent hydrogen atom (1.77 \AA~with CCSD, 1.75 \AA~with SA-CASSCF). 

For azobenzene, we focus on an S$_0$/S$_1$ MECI (CI-rot) involved in 
the photoinduced cis-trans isomerization reaction~\cite{yu2015probing}.
In Fig.~\ref{fig:azobenzene-comparison}, we compare the CCSD $\varepsilon$-MECI that is converged with $\varepsilon$=0.20 eV with the SA-CASSCF structure from Ref.\citenum{yu2015probing}, using a 6-31G basis set. In the isomerization pathways, 
the most important internal coordinates are the dihedral angles C$_2$N$_1$N$_2$C$_3$, C$_1$C$_2$N$_1$N$_2$ and N$_1$N$_2$C$_3$C$_4$. 
All these dihedrals are well described with CCSD, and the largest deviation from the SA-CASSCF structure is only 5$\degree$ and occurs for the C$_1$C$_2$N$_1$N$_2$  angle (175$\degree$ in CCSD and 180$\degree$ in SA-CASSCF). 
Overall, the CCSD $\varepsilon$-MECI is in good agreement with the reference structure, with only small differences in the internal coordinates not directly
involved in the cis-trans isomerization (see Supporting Information). 

Finally, we consider two S$_0$/S$_1$ MECIs for uracil. The first one is characterized by a puckering of the C$_5$ atom and consequently out-of-plane bending of H$_5$, while the second MECI involves an out-of-plane bending of the O$_8$ atom (see Fig.~\ref{fig:uracil-comparison}). 
Both structures have previously been identified at various levels of theory.\cite{zhang2014excited,nachtigallova2011nonadiabatic,merchan2006unified,lan2009photoinduced,matsika2004radiationless} We will compare our CCSD/cc-pVDZ structure with spin-flip TDDFT (SF-TDDFT)/6-31+G~\cite{zhang2014excited} and CASSCF(10/8)/6-31G*~\cite{nachtigallova2011nonadiabatic}.
Following the nomenclature suggested by Nachtigallová et al.~\cite{nachtigallova2011nonadiabatic}, we  
refer to the
two structures as $^6$S$_5$ and \textit{oop}-O, respectively.

In the $^6$S$_5$ structure, the C$_5$-C$_6$ double bond assumes an ethylene-like structure, with the dihedral C$_4$C$_5$C$_6$H$_5$ describing the bending of the H$_5$ atom. 
This angle is 114$\degree$ with CCSD, and is in line with the CASSCF value of 112$\degree$\cite{nachtigallova2011nonadiabatic}. 
The C$_5$ puckering is described by the C$_4$C$_5$C$_6$N$_1$ dihedral angle. In this case, the three selected methods agree, with 53$\degree$ with SF-TDDFT\cite{zhang2014excited} and 52$\degree$ for CCSD and CASSCF.\cite{nachtigallova2011nonadiabatic} 
Internal coordinates describing the overall structure of the ring are also found to be in agreement among the different methods (see Supporting Information).
In the \textit{oop}-O structure, the out-of-plane bending of the O$_8$ atom is described by the dihedral angle C$_2$N$_3$C$_4$O$_8$, we have 94$\degree$ in CCSD and 113$\degree$ in CASSCF.
Compared to the $^6$S$_5$ structure, the ring is now planar, as described by the C$_4$C$_5$C$_6$N$_1$ dihedral angle, which is 0.5$\degree$ in CCSD and 1.0$\degree$ in CASSCF.
Similar agreement is found for the other internal coordinates, which describe an overall similar molecular geometry with the two methods, with the main difference being in the C$_4$-O$_8$ bond (0.10 Å). 

In this Letter, we have examined a simple algorithm for locating minimum energy conical intersections, showing that it
converges to a minimum energy structure on a hypersurface where the energy difference between the states is $\varepsilon$ and that it provides accurate
structures for sufficiently small $\varepsilon$. Moreover, it does not require the evaluation of nonadiabatic coupling vectors, which are often not available in programs and are sometimes not easily available for a given electronic structure method. 
By enforcing the energy difference between the states to be 
small but non-zero, the algorithm can be used to avoid the numerical artifacts and convergence issues of coupled cluster theory in the vicinity of conical intersections. This has allowed us to investigate $\varepsilon$-MECI structures between the ground and first excited states in coupled cluster theory, which has not been possible before. Our results show that CCSD, despite its convergence issues, can provide an accurate description of conical intersections with the ground state, agreeing quantitatively with other state-of-the-art methods. This suggests that coupled cluster theory may be a good candidate for nonadiabatic dynamics simulations targeting non-radiative relaxation to the ground state. 

\section*{Author contributions}
SA and EFK conceived the project and performed calculations. HK supervised the project. SA and HK analyzed the results and wrote the first draft of the manuscript. All authors discussed and revised the manuscript.

\begin{acknowledgement}
We thank Marcus T. Lexander and Federico Rossi for enlightening discussions. This work was supported by the European Research Council (ERC) under the European Union's Horizon 2020 Research and Innovation Program (grant agreement No. 101020016).
\end{acknowledgement}

\begin{suppinfo}
Study of the convergence of the algorithm, comparison of CCSD and SCCSD S$_1$/S$_2$ $\varepsilon$-MECIs for uracil, and additional data for the S$_0$/S$_1$ $\varepsilon$-MECIs for uracil and azobenzene. 
\end{suppinfo}

\bibliography{bibliography}

\includepdf[pages={{},-}, pagecommand={\thispagestyle{empty}}]{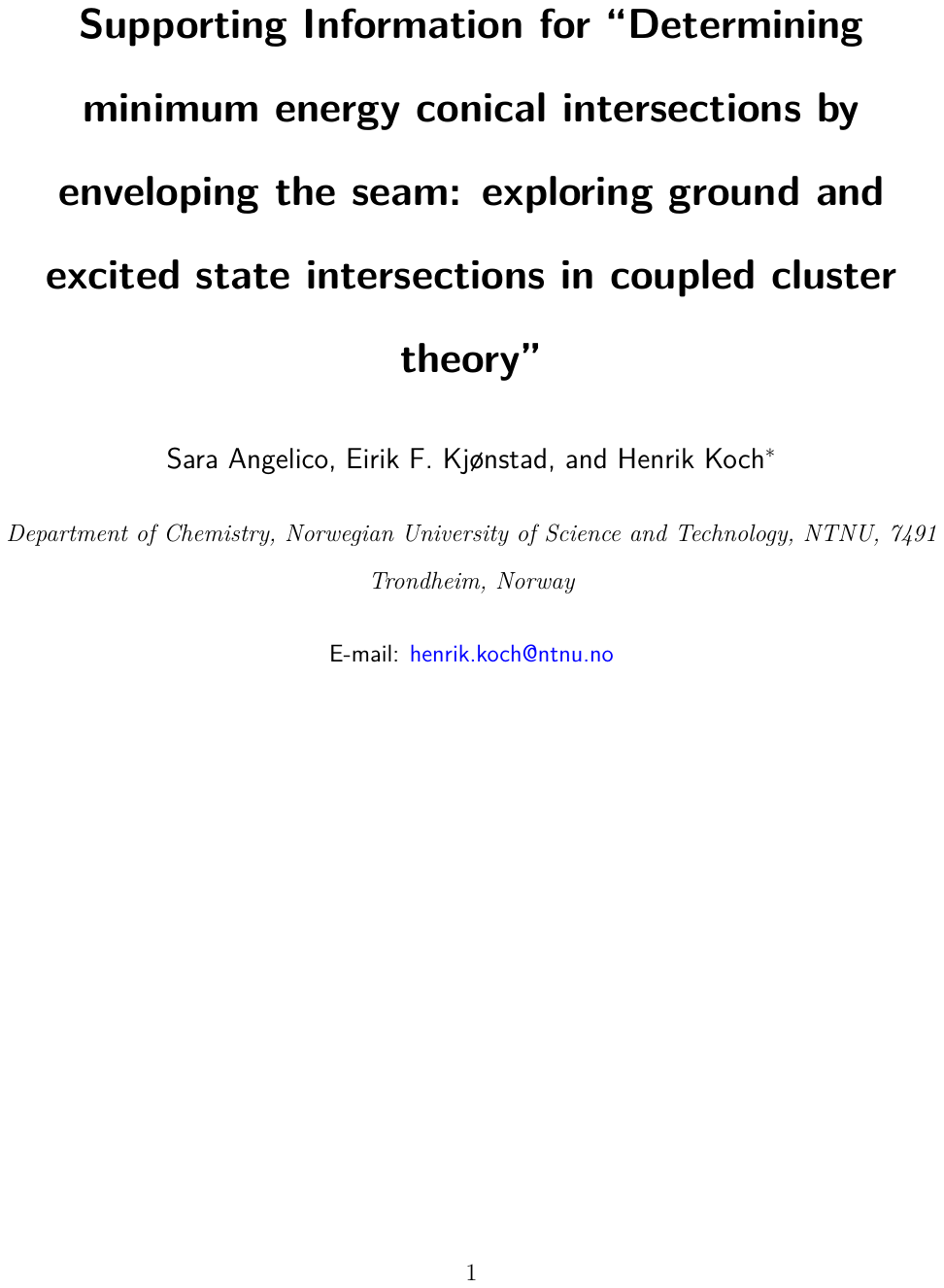}

\end{document}